\documentclass[preprint]{elsarticle}

\usepackage{lineno}
\usepackage{multirow}
\usepackage{amsmath}
\usepackage[figuresright]{rotating}
\usepackage{float}
\usepackage{array}

\journal{Journal of \LaTeX\ Templates}

\biboptions{authoryear}

\begin{document}

\begin{frontmatter}

\title{Disarranged Zone Learning (DZL): An unsupervised and dynamic automatic stenosis recognition methodology based on coronary angiography}

\author{Yanan Dai\fnref{myfootnote}}
\address{School of Management, Fudan University. No.220 Handan Road, Shanghai, China}
\fntext[myfootnote]{The authors contributed equally to this work.}

\author{Pengxiong Zhu\fnref{myfootnote}}
\address{Department of Cardiac Surgery, Shanghai East Hospital, Tongji University. No.150, Jimo Road, Shanghai, China}

\author{Bangde Xue}
\address{Department of Cardiac Surgery, Shanghai East Hospital, Tongji University. No.150, Jimo Road, Shanghai, China}

\author{Yun Ling}
\address{Department of Cardiac Surgery, Shanghai East Hospital, Tongji University. No.150, Jimo Road, Shanghai, China}

\author{Xibao Shi}
\address{Department of Cardiovascular Medicine, Ruijin Hospital, Shanghai Jiao Tong University School of Medicine. No.197, Ruijin er Road, Shanghai, China}

\author{Liang Geng}
\address{Department of Cardiology, Shanghai East Hospital, Tongji University. No.150, Jimo Road, Shanghai, China}

\cortext[mycorrespondingauthor]{Corresponding author email address: liujun52@126.com, zhangqnh@hotmail.com}

\author{Qi Zhang\corref{mycorrespondingauthor}}
\address{Department of Cardiology, Shanghai East Hospital, Tongji University. No.150, Jimo Road, Shanghai, China}

\author{Jun Liu\corref{mycorrespondingauthor}}
\address{Department of Cardiac Surgery, Shanghai East Hospital, Tongji University. No.150, Jimo Road, Shanghai, China}

\author{}




\par

\begin{abstract}
We proposed a novel unsupervised methodology named Disarranged Zone Learning (DZL) to automatically recognize stenosis in coronary angiography. The methodology firstly disarranges the frames in a video, secondly it generates an effective zone and lastly trains an encoder-decoder GRU model to learn the capability to recover disarranged frames. The breakthrough of our study is to discover and validate the Sequence Intensity (Recover Difficulty) is a measure of Coronary Artery Stenosis Status. Hence, the prediction accuracy of DZL is used as an approximator of coronary stenosis indicator. DZL is an unsupervised methodology and no label engineering effort is needed, the sub GRU model in DZL works as a self-supervised approach.  So DZL could theoretically utilize infinitely huge amounts of coronary angiographies to learn and improve performance without laborious data labeling. There is no data preprocessing precondition to run DZL as it dynamically utilizes the whole video, hence it is easy to be implemented and generalized to overcome the data heterogeneity of coronary angiography. The overall average precision score achieves 0.93, AUC achieves 0.8 for this pure methodology. The highest segmented average precision score is 0.98 and the highest segmented AUC is 0.87 for coronary occlusion indicator. Finally, we developed a software demo to implement DZL methodology.

\end{abstract}


\end{frontmatter}


\section{Introduction}

Deep learning has achieved great success in computer vision, among which in the field of medical, the first application of deep learning is medical image processing. For example, magnetic resonance images of the brain, ultrasound images of breast nodules, CT and X-ray images of the lungs, these images reflect the structure of organs and the attributes of lesions through continuous groups of pictures from different angles.\citep{Miotto2018} Each image is a static reflection of the lesion under a certain angle. In cardiac imaging research, deep learning is applied to the segmentation of LV and CCS based on CMR, CT, ultrasound images and to the extraction of coronary artery and its centerlines mainly based on CCTA. However, as far as we know, discussion of deep learning in coronary angiography which is regarded as the golden rule for coronary artery disease diagnostics is very limited, mainly for two reasons: 

1)	Coronary angiography(CAG) is a video, which records the dynamic process of contrast agent flowing in the blood vessel. For the part that is not able to be developed by contrast agent, it is the location of suspected lesions. The methodology of reflecting the lesions causes the analysis of coronary angiography extended to the three-dimensional space, by adding the extra dimension of time. For each lesion, adjacent images are no longer independent. The strong dependence makes the analysis of coronary angiography be based on the whole of the video. The correlation between adjacent images makes the frame sequence unable to maintain Markov property, which further increases the complexity of modeling. There are several studies of deep learning based on coronary angiography, but they only select the certain frame from the angiogram to do the further analysis. This is a static assessment of the lesion and can only get anatomical feature. After all, anatomic stenosis does not necessarily mean true ischemia. Our methodology is based on dynamic changes in the lesion assessment, the results are from an functional point of view. Results from functional method are more correlated with ischemia.\citep{Kumamaru2020}Therefore, the research difficulty of deep learning in coronary angiography, especially in a dynamic perspective, is higher than that in other fields while such research results are more beneficial. 

2)	The annotation of medical images needs to consume huge medical resources, especially where such annotation can only be completed by professional doctors. Coronary angiography is a video, each has several frames, so the annotation will consume even more. This could be the reason that the previous studies, even based on coronary angiography, only select key frames from the whole video to annotate and analysis. Such static methods lose the tremendous information advantage of coronary angiography. Other related studies just use CCTA instead of CAG and are constrained to the very small sample size probably due to the huge consumption of labeling engineering. Our proposed methodology is totally label-free and overcomes the laborious difficulty. Therefore it is very easy to be implemented and generalized. 

In this paper, a methodology (DZL) is originally designed to overcome the two research bottlenecks mentioned above. Sequence Intensity is defined and used as an indicator for evaluating the stenosis. In order to establish the link between the two, we creatively define Reorder Difficulty to measure the Sequence Intensity. That is to say, the stronger the Sequence Intensity is, the easier it is to be restored after the sequence is disrupted, and the lower the Reorder Difficulty is. Then a recurrent neural network(RNN) model is designed to imitate human's capability to recover the disordered frames sequence. The model prediction accuracy is creatively used as an approximator for Reorder Difficulty. The higher the model accuracy is, the lower Reorder Difficulty is. Finally, the relationship between Sequence Intensity and coronary artery stenosis status is discovered and validated via using the approximator measure of Sequence Intensity. 

To sum up, our study is the first to deliver the definition of Sequence Intensity and Reorder Difficulty of coronary angiography, also we are the first to approximate Reorder Difficulty using deep learning model prediction accuracy. Last but not least, we are the first to discover and validate the relationship between Sequence Intensity and Coronary Artery Stenosis Status. Amazingly, the overall process even does not need any manual labeling. 

We evaluate the contributions of the proposed methodology from the two major challenges of coronary angiography data, that is, how the methodology overcomes the difficulties in these two dimensions: 

Time dependency: coronary angiography is a video, there is a strong dependence between frames, if we use the conventional modeling method, since the lack of Markov property and the model will be extremely complex and hard to generalize. The novel methodology designed by this paper makes full use of the time dependence between frames and essentially establishes the relationship between the time dependence and the lesion status. Time dependence is no longer a challenge to the methodology, but has become the core principle of the methodology, which is an idea of turning waste into treasure. 

Heterogeneity: coronary angiography is patient based. Coronary angiography from different patients show different dynamic process in terms of speed, position, vessels thickness, heart size, heart rate, spinal occlusion status which are hard to generalize, especially in dynamic perspective. The methodology captures the implicit relationship between the Sequence Intensity and Coronary Artery Stenosis Status, this is a high-level intrinsic relationship regardless of left or right dominance, proximal occlusion or distal stenosis, size, shape and rate of heart, error from instruments etc.. Also, it is totally label-free. Hence the methodology is possible to be generalized in various videos of different patients with strong robustness and cross-platform migration capability. 

The overall system architecture of the methodology is shown in Figure \ref{system_architecture}: 

\begin{figure}[H]
\centering
\includegraphics[scale=0.45]{"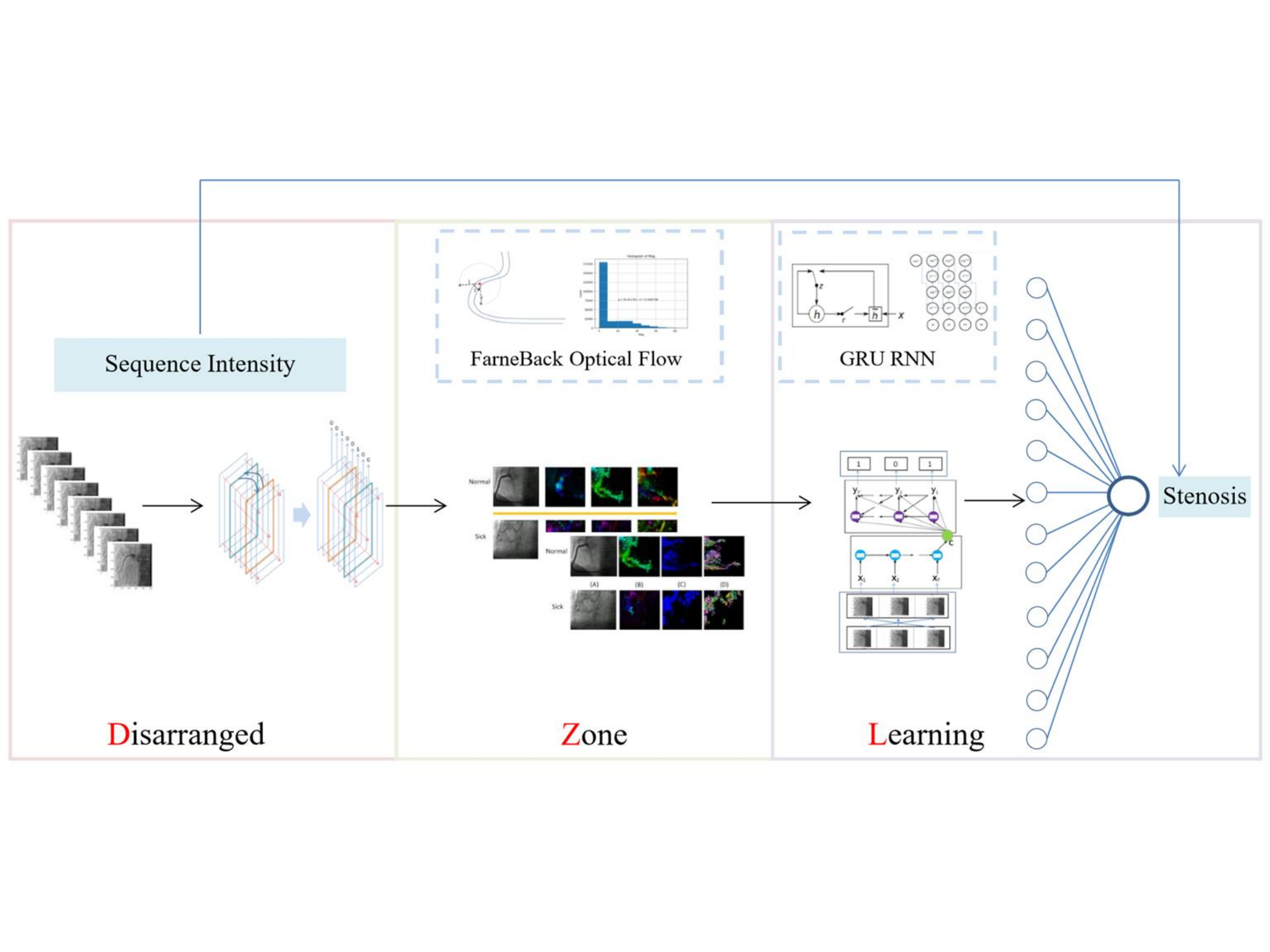"}
\caption{\textbf{Overall System Architecture.} The overall methodology can be divided into three parts: disarranged, zone and learning. The disarranged part is functional to disorder the frames of angiography segment. The zone part is functional to generate an effective optical flow zone to capture the sequence information of segments. The learning part is to learn how to recover the disarranged order accordingly. }
\label{system_architecture}
\end{figure}

The sections are organized as follows. Section 1 is the overall introduction of our study including background introduction and main contributions. Section 2 is the theory development. Section 3 is the methodology specification. Section 4 is the analysis results. Section 5 is the discussion about advantage, extension and potential of this methodology. Section 6 is the limitations to further provide a future work direction.  

\section{Theory}\label{theorem}

In this section, we proposed the main theorem of DZL.

\newdefinition{dfn}{Definition}[section]
\newtheorem{lem}{Lemma}[section]
\newproof{pot}{Proof of Lemma 2.1}
\newtheorem{thm}{Theorm}[section]
\newtheorem{prop}{Proposition}[section]

\begin{dfn}
\emph{Sequence Intensity}. It measures the correlation strength of the adjacent frames in a video. It is defined as the cumulative counts if the true next frame is with biggest probability regarded as the corresponding next frame given its previous frames for all the frames.
\begin{equation}
Tindex_t=
\begin{cases}
1& \text{\emph{t} = $\mathop{\arg\max}\limits_{t'}$}\frac{P{ \left( {F_{t'} \left| F_{t-1},F_{t-2}...,F_{0}\right. } \right) }}{(\sum_{t'^-}P{ \left( {F_{t'^-} \left| F_{t-1},F_{t-2}...,F_{0}\right. } \right) })+P{ \left( {F_{t'} \left| F_{t-1},F_{t-2}...,F_{0}\right. } \right) }}\\
0& \text{Otherwise}
\end{cases}
\end{equation}

\begin{equation}
SequenceIntensity = \sum_{t}Tindex_t
\end{equation}

\end{dfn}

\begin{dfn}
\emph{Reorder Difficulty}. It measures the correlation strength of the adjacent frames in a sequence-shuffled video. It is defined as the cumulative counts if the shuffled next frame is with biggest probability regarded as the corresponding next frame given its previous frames for all the frames.
\begin{equation}
Findex_s=
\begin{cases}
1& \text{\emph{s} = $\mathop{\arg\max}\limits_{s'}$}\frac{P{ \left( {F_{s'} \left| F_{s-1},F_{s-2}...,F_{0}\right. } \right) }}{(\sum_{s'^-}P{ \left( {F_{s'^-} \left| F_{s-1},F_{s-2}...,F_{0}\right. } \right) })+P{ \left( {F_{s'} \left| F_{s-1},F_{s-2}...,F_{0}\right. } \right) }}\\
0& \text{Otherwise}
\end{cases}
\end{equation}

\begin{equation}
ReorderDifficulty = \sum_{s}Findex_s
\end{equation}

\end{dfn}

From the above definitions, Sequence Intensity and Reorder Difficulty are two negative correlated variables and we have the following relationship: 

\begin{lem}\label{lem}
\emph{Complementary Order Index}.
\begin{equation}
SequenceIntensity = T - \alpha - ReorderDifficult
\end{equation}
\end{lem}
In which, T is the total frames number in a video and $\alpha$ is the counts of the positions that are not shuffled during the random shuffling process. 

\begin{pot}
For the position $t$ that is shuffled, if $Tindex_t = 1$, then we must have: $Findex_t = 0$ and vise versa. $Tindex_t = 1$ means the true next frame $F_t$ is regarded as the next frame with biggest probability while $Findex_s = 1$ means the shuffled next frame $F_s$ is regarded as the next frame with biggest probability. The two statements is exclusive when the position t is shuffled. For the position that is not shuffled, then we have $Tindex_t = Findex_t$ correspondingly. 
\end{pot}

The contrast agent injection process of a normal coronary artery is a clear time-sequential dynamic process and each frame is an evolutionary step towards the dynamic process. The abnormal coronary artery, due to vascular stenosis or occlusion which affects the smooth injection of contrast agent, results in the dynamic process more blurred and not easy to be distinguished. We originally discover the Sequence Intensity as an indicator of Coronary Artery Stenosis Status in terms of normal/abnormal. And Theorem\ref{thm} is established. 

\begin{thm}\label{thm}
Sequence Intensity of coronary angiography is an indicator of Coronary Artery Stenosis Status. 
\end{thm}

Stenosis and occlusion will hinder the dynamic contrast agent injection process and make some adjacent frames with less sequence information. Hence, two propositions are proposed as follows: 

\begin{prop}
Sequence Intensity of coronary angiography is negatively correlated with Coronary Artery Stenosis Status. The higher probability there is stenosis with, the weaker the Sequence Intensity is.
\end{prop}

From Lemma\ref{lem}, we have the following proposition: 

\begin{prop}
Reorder Difficulty of coronary angiography is positively correlated with Coronary Artery Stenosis Status. The higher probability there is stenosis with, the greater the Reorder Difficulty is.
\end{prop}

\section{Methodology}

\begin{figure}[H]
\centering
\includegraphics[scale=0.45]{"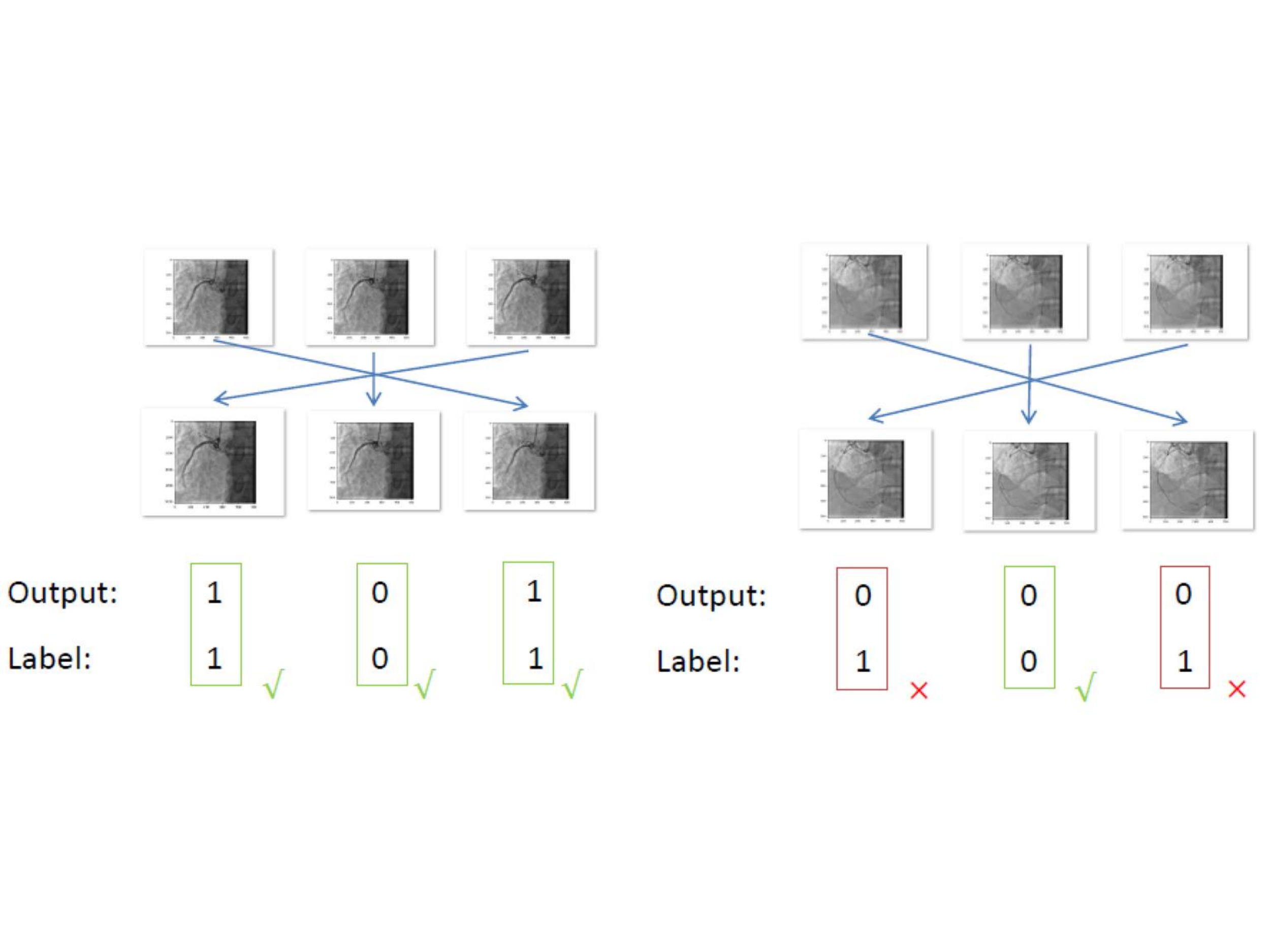"}
\caption{\textbf{The Intuition.} The sub-left figure is the illustration of a sequence of frames from a video of normal coronary vessels. For example, the first and third frames are exchanged and the second frame is not moved. Hence, such disarranging cause the first and third frames labeled as 1 and the second frame labeled as 0. For the sub-right figure, the settings are the same except that it is an abnormal and occluded right coronary artery. We argue the normal coronary angiography is with higher Sequence Intensity and high model prediction accuracy. The example shows that the sub-left output-label pairs are consistent, indicating a high model prediction accuracy. Meanwhile, the sub-right output-label pairs are different, indicating a comparatively low model prediction accuracy. The methodology is developed in details in the subsequent subsections.}
\label{Disarranged}
\end{figure}

The contrast agent injection process of a normal coronary artery is a clear time-sequential dynamic process and each frame is an evolutionary step towards the dynamic process. The abnormal coronary artery, due to vascular stenosis or occlusion which affects the smooth injection of contrast agent, results in the dynamic process more blurred and not easy to be distinguished. Hence we argue the Sequence Intensity, defined in Section \ref{theorem} could be an indicator of Coronary Artery Stenosis Status in terms of normal/abnormal. The intuition of the methodology is illustrated in Figure \ref{Disarranged}

\subsection{Step 1: Disarranging Sequence Order of Coronary Angiography}

A coronary angiography video is composed of many frames and each frame is an image which records development state of the vessels at a certain time when the contrast agent is injected. We first shuffle the frames sequence of the video. We set $\theta$, the shuffling percentage, equal to 50\%. Half of the frames in a video is shuffled while the other half is not shuffled. If the total number of frames of the video is odd, we round the half to be shuffled. 

\subsection{Step 2: Zone Generating}

We are concerned about the dynamic imaging process of blood vessels during contrast injection. In order to overcome the difficulty of high dimension, we use optical flow technology to extract sequence features and form an effective zone to calculate on. 

\begin{figure}[H]
\centering
\includegraphics[scale=0.45]{"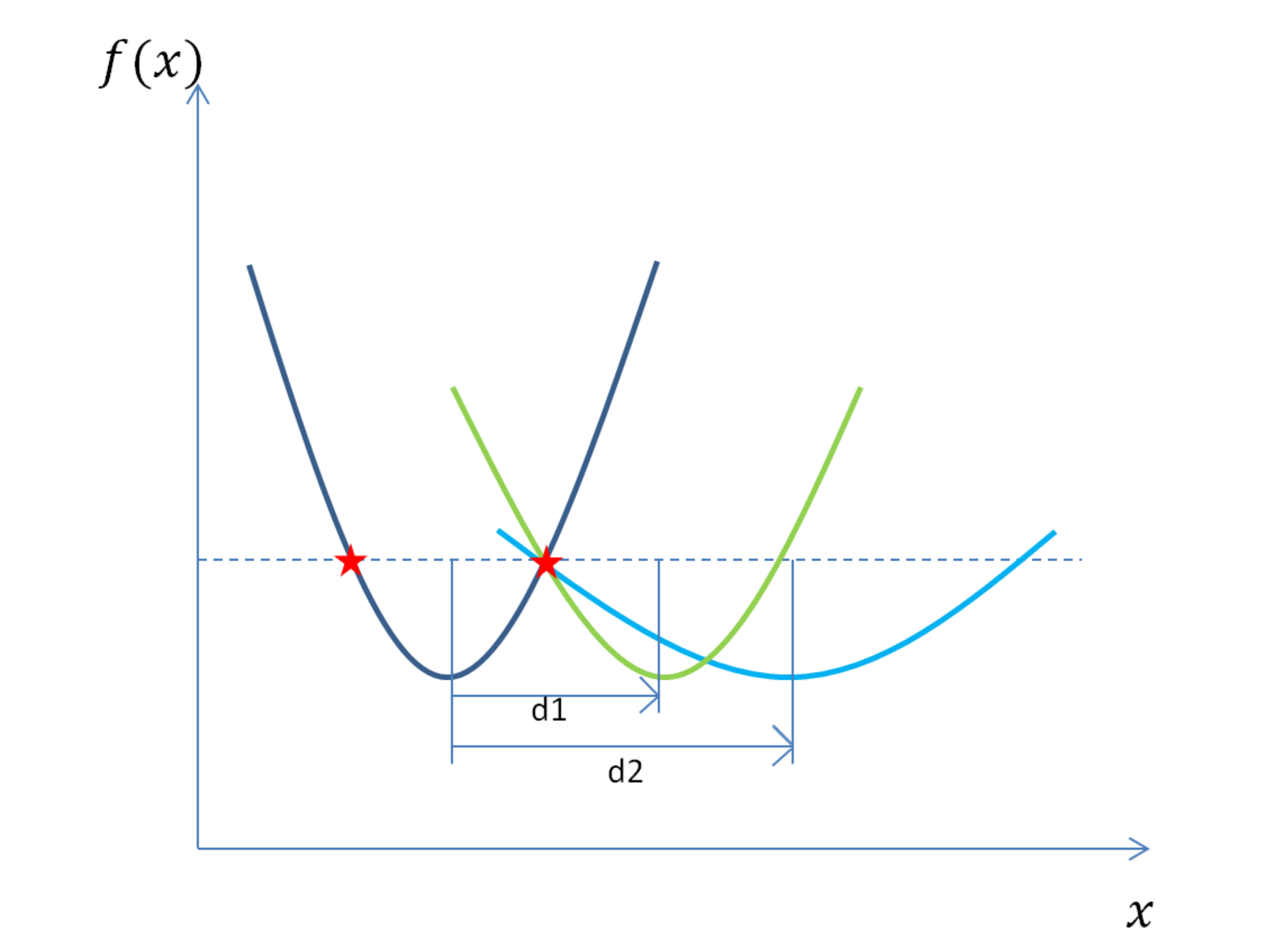"}
\caption{\textbf{2-dimension Farneback Illustration.} The principle of dense optical flow is to find the optimal flow direction and magnitude through the polynomial expansion in the pixel field based on the assumption of pixel displacement invariance. We reduce the high dimension description in Farneback into 2 dimensional conditions. The x-axis in the figure is the spatial position of the pixel point, f(x) is the corresponding second polynomial expansion of pixel value. Based on the assumption of displacement invariance, the movement of the pixel is illustrated as moving from the left star to the right star, keeping f(x) as the same, and the polynomial expansion can be changed from the dark blue quadratic curve to either the light green quadratic curve or the light blue quadratic curve or other quadratic curve passing through the second star. However, different quadratic curve (second polynomial expansion) corresponds to different optical flow direction and magnitude, the objective is to find the optimal most satisfies the displacement invariance constraint. }
\label{farneback}
\end{figure}

Optical flow is the instantaneous motion speed of the pixels of a spatially moving object in the observed imaging plan. Optical flow uses the change of pixels in the image sequence in the time domain and the correlation between adjacent frames to find the corresponding relationship between frames, so as to calculate the motion information of objects between adjacent frames. Optical flow can be generally divided into two categories: sparse optical flow and dense optical flow. \citet{Lucas1981} is a typical method to calculate the sparse optical flow. This method needs to use some methodology to locate the corner points as the initial points of the optical flow. We used \citet{Shi1994} method to locate the corner points but find these points are less related to vessels due to various disturbances such as spinal pixels, heartbeat, angle transformation etc. That is, the corner points are not generated on the vessel, so the generated optical flow fails to represent the dynamic effect of the contrast agent injection process. So we chose to compute the dense optical flow proposed by \citet{Farneback2003}. It is to compute the optical flow of all pixels between frames. It is mainly through polynomial expansion and the assumption of pixel displacement invariance to reach the optimal optical flow. Figure \ref{farneback} is a representation of the Farneback dense optical flow mechanism in 2 dimensions perspective. 

The Farenback optical flow calculation is based on \eqref{1},\eqref{2}. \eqref{1} is the second polynomial expansion of pixel point x. \eqref{2} is the second polynomial expansion of pixel point x plus a displacement. According to the displacement variance, $f_1 (x)=f_2 (x-d)$, and we have coefficient constraints \eqref{3}$-$\eqref{6}:

\begin{equation}\label{1}
f_{1}(x) = a_{1}x^{2}+b_{1}x+c_1
\end{equation}

\begin{equation}\label{2}
f_{2}(x-d) = a_{2}(x-d)^{2}+b_{2}(x-d)+c_1
\end{equation}

\begin{equation}\label{3}
a_1 = a_2
\end{equation}

\begin{equation}\label{4}
b_1 = b_2-2a_{2}d
\end{equation}

\begin{equation}\label{5}
c_1 = a_{2}d^{2}-b_{2}d+c_2
\end{equation}

\begin{equation}\label{6}
\triangle b=-\frac{1}{2}(b_1+b_2)
\end{equation}

Our objective is to select the optimal optical flow that satisfies the above constraints most.  

We then encounter the problem of further selection of effective feature points and form an active zone to enforce the disarranging function on. Because dense optical flow calculates the optical flow of all pixels in a frame, only a small part of which is related to blood vessels. If all the pixels are analyzed, a compensation mechanism will work which puts a negative impact on the analysis of the results. Since coronary angiography also records the movement of part of the spine, heartbeat as well as some movement of the camera angles. Such additional dynamics are easier to generate strong Sequence Intensity, so it may make it easier to recover the disorder via using such additional information though the vascular visualization part is not easy to be recognized. Bias may correspondingly be generated. 

Hence, it is necessary to develop a screening mechanism to select an “effective zone” from where the optical flows are generated from the pixels of vessel. 

\begin{figure}[H]
\centering
\includegraphics[scale=0.45]{"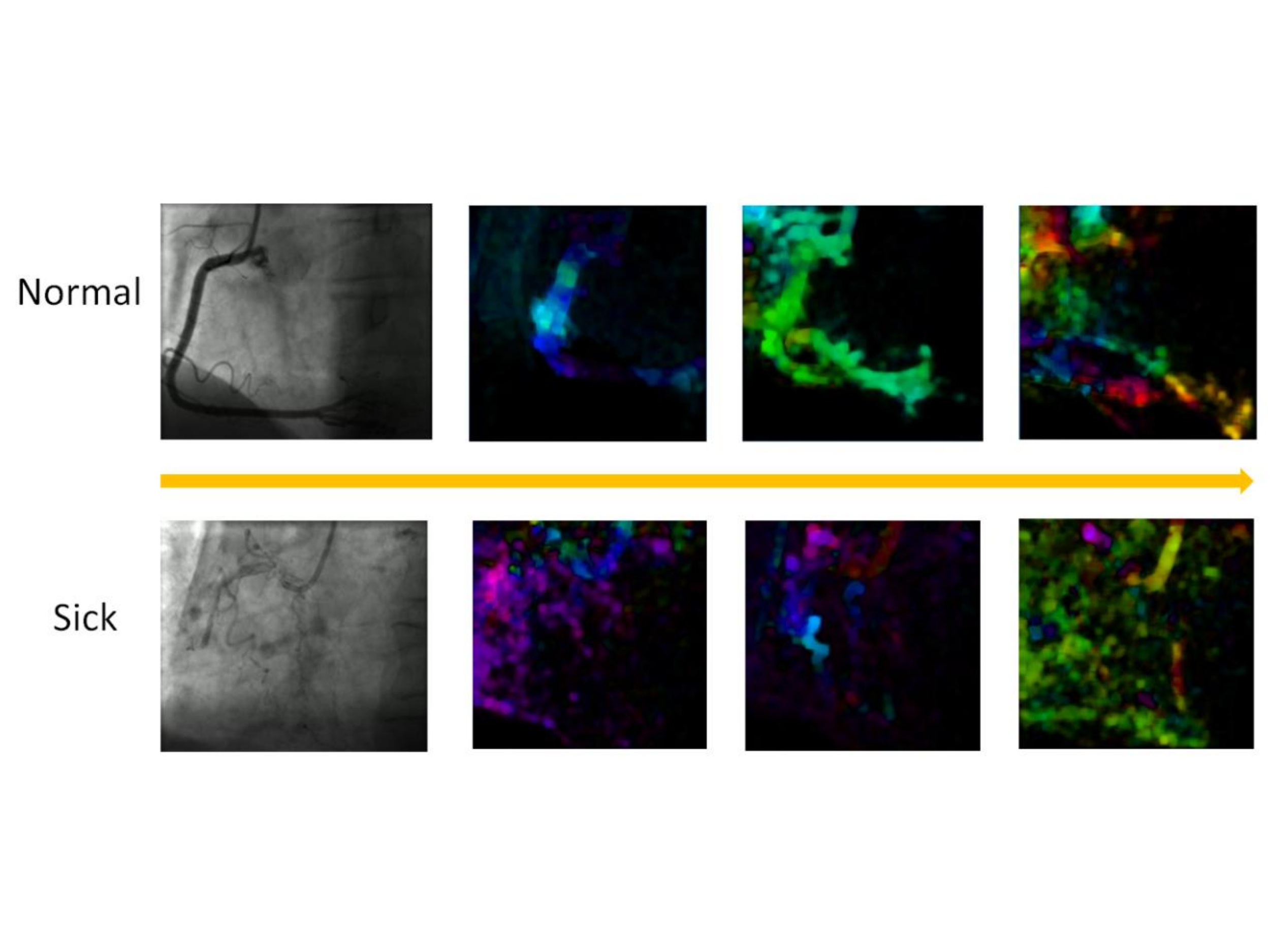"}
\caption{\textbf{Dense Optical Flow Calculation Process.} The first row is from a normal coronary angiography, the second row is from an abnormal coronary. The orange arrow is the time axis. The dense optical flow is developed along the time axis. }
\label{Densecalc}
\end{figure}

Figure\ref{Densecalc} is a dense optical flow calculation process, from which we get such a view that the points associated with blood vessels generally change sharply in color and brightness. From the visualization process, different color and brightness express different magnitudes and angles of optical flow. 

\begin{figure}[H]
\centering
\includegraphics[scale=0.45]{"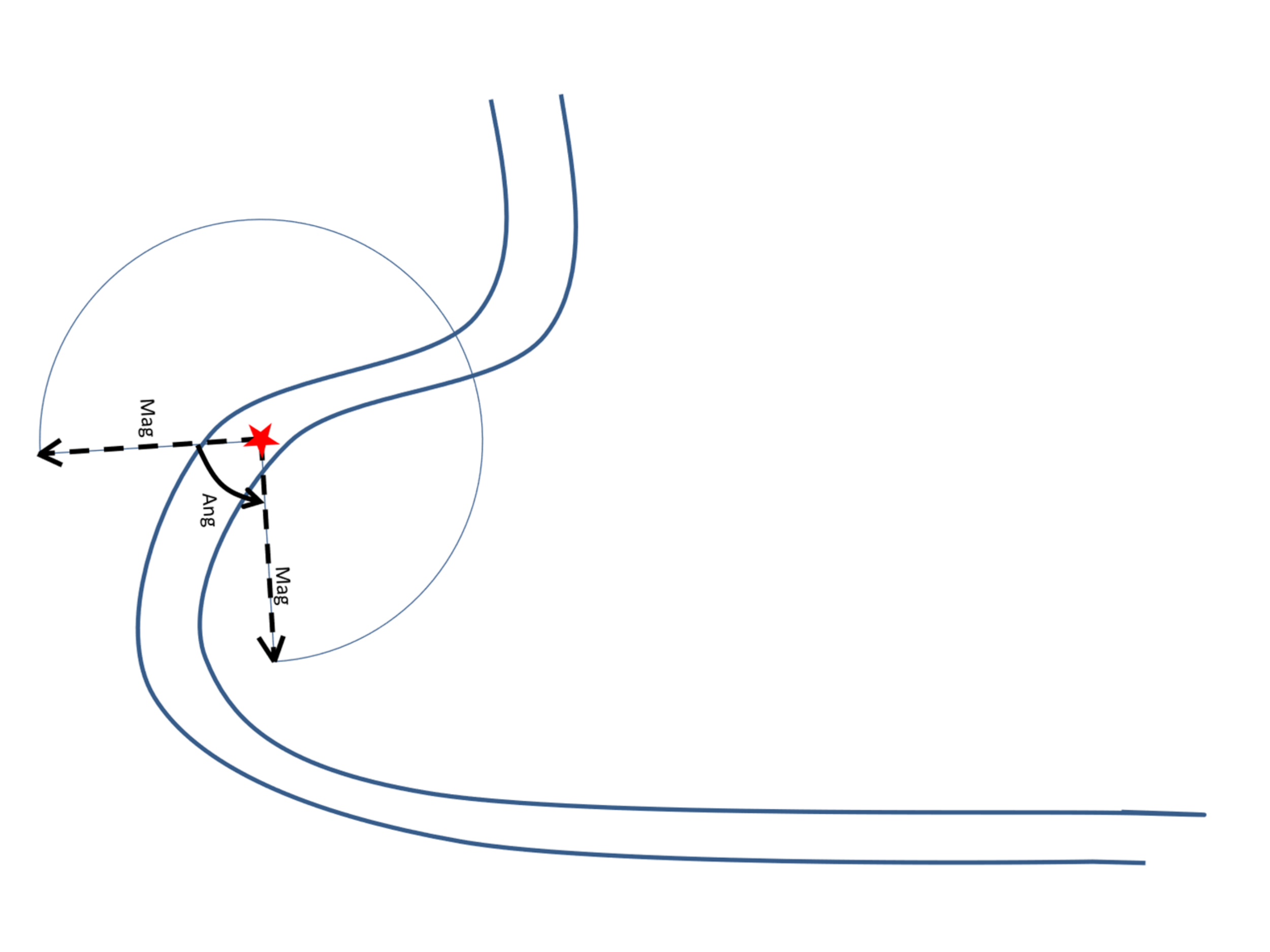"}
\caption{\textbf{Polar Coordinate Plane View}}
\label{polar}
\end{figure}

\begin{figure}[H]
\centering
\includegraphics[scale=0.45]{"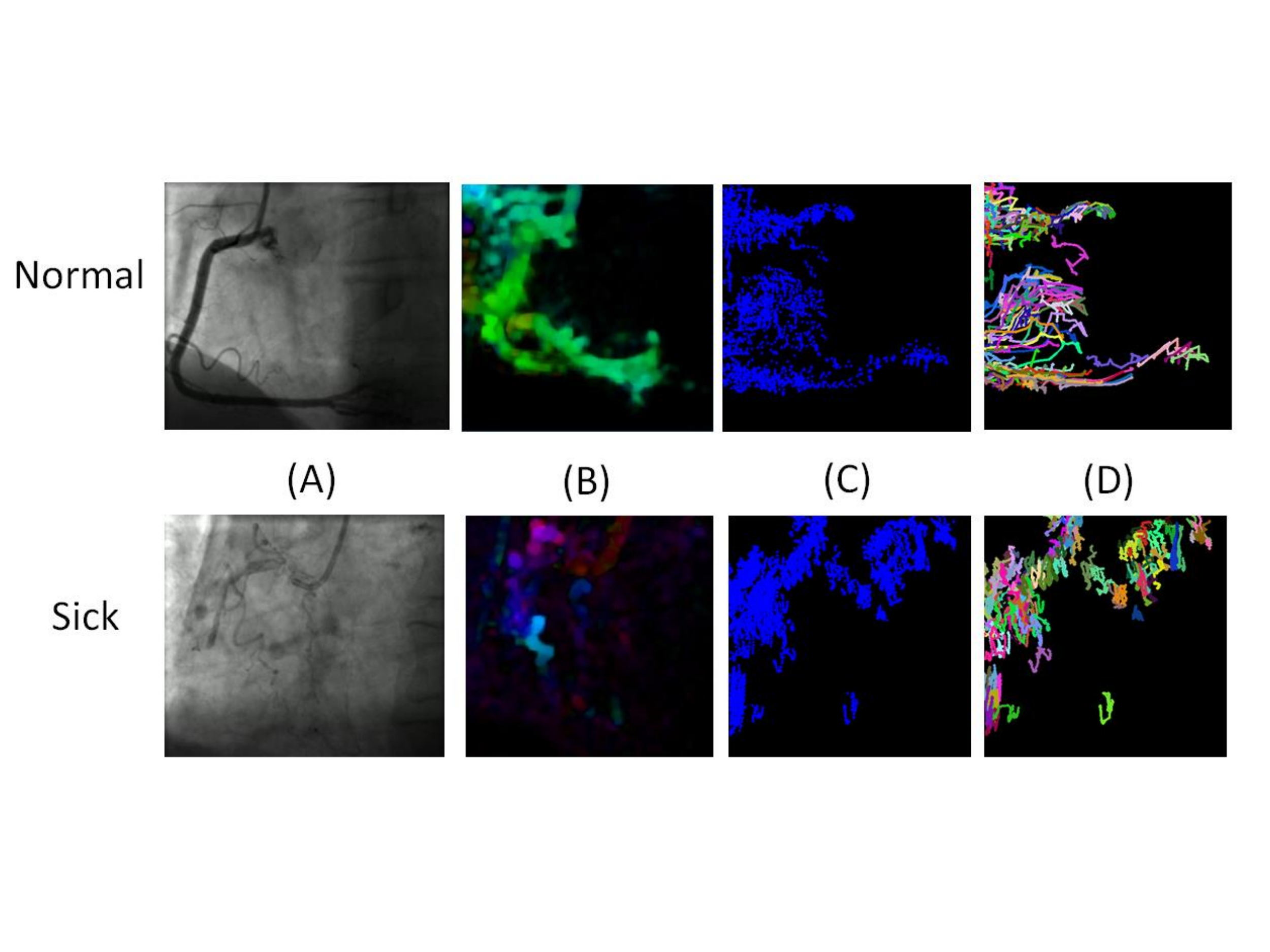"}
\caption{\textbf{Effective Zone Generation.}The first row is from a normal coronary angiography, the second row is from an abnormal coronary. Column (A) is from the original angiography. Column (B) is the calculation results of Farneback optical flow of a frame in the segment. Column (C) is the randomly selected 100 optical flow points from the optical flow point sets above 80\% percentile of magnitude variance. Column (D) is the optical flow trace from 100 points selected. }
\label{trace}
\end{figure}

Figure\ref{polar} is a vessel in polar coordinate plane. We finally select the point whose flow magnitude variance is above 80 percentile as effective. Moreover, we aggressively and randomly select 100 points from the 52428 effective points as the final zone. A final zone and its optical flow trace are illustrated in Figure \ref{trace}. The above hyperparameters are selected based on try and error. 

\subsection{Step 3: Learning}\label{Step 3: Learning}

The objective of the learning model is to be trained so as to learn the capability to recover disordered frames like human. To simplify the model, we design the model as a two-class classification model, that is, we train the model to tell if the specific frame is disarranged or not. The labels guiding the training process are generated automatically comparing to the original order. Once the model is trained to learn the recovering capability, its accuracy is used as an approximate to Reorder Difficulty and therefore be an indicator of Coronary Artery Stenosis Status. The model training process doesn't require any label engineering, in other words, it is not necessary to know the "answer" of Coronary Artery Stenosis Status beforehand, the model intrinsically works in an unsupervised learning approach. 

\begin{figure}[H]
\centering
\includegraphics[scale=0.45]{"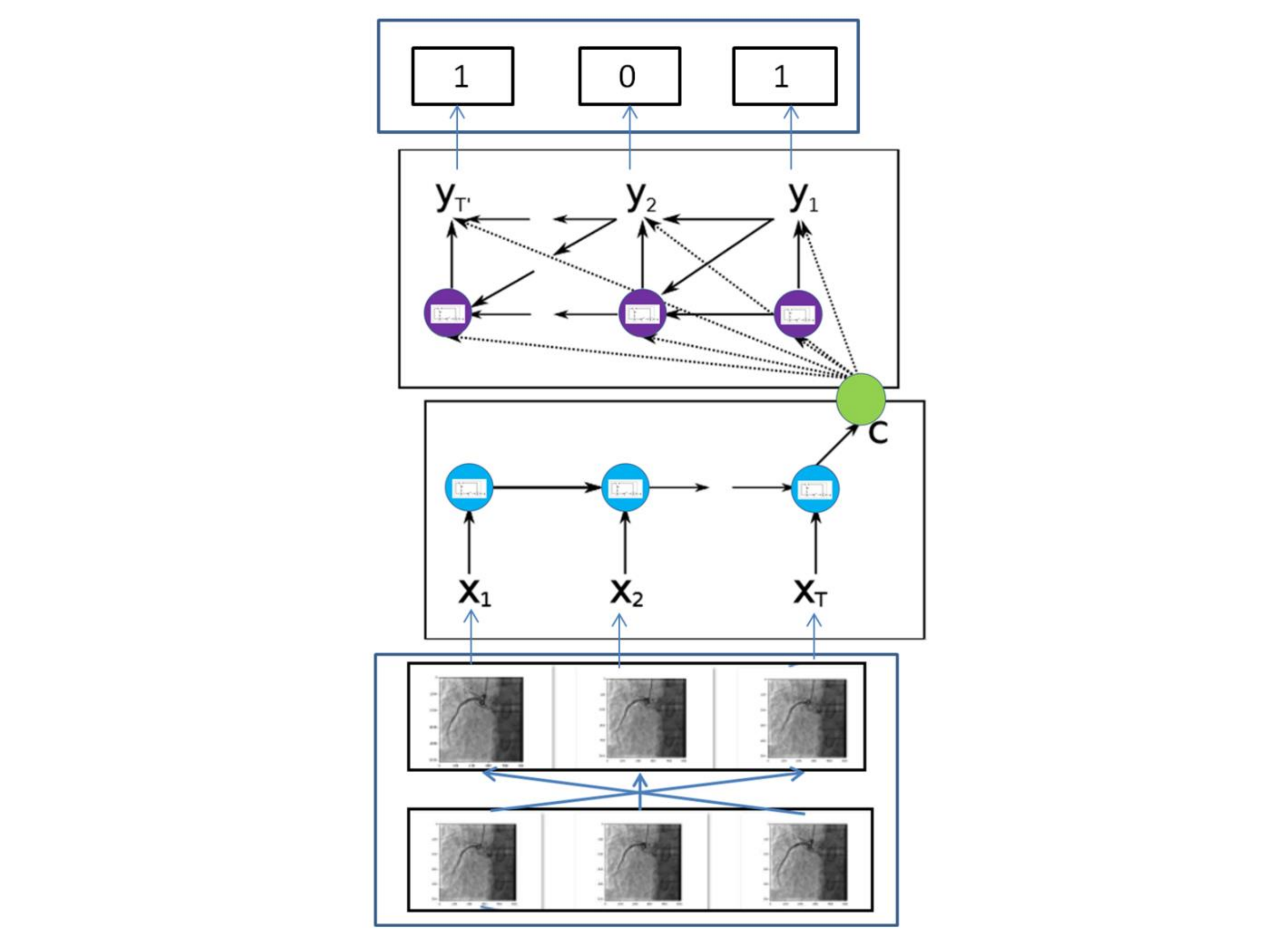"}
\caption{\textbf{Encoder-Decoder GRU Model.}The bottom block is effective zone of disarranged frames as the input of the Encoder-Decoder GRU model. The X block is the encoder module, each node is a GRU unit. The green circle C is the fixed length vector output by the encoder. The y block is the decoder module, each node is also a GRU unit. The loss of the model is calculated through the output of the Y block and the binary label generated automatically via the disarranging status of the input frames. If the specific frame is disarranged, it is correspondingly labeled as 1, otherwise, labeled as 0.}
\label{RNN}
\end{figure}

In terms of the learning of recovery capability, as our input is the sequence of disarranged optical flow from the final effective zone, we chose the recurrent neural network (RNN) which is specifically used to learn and predict the sequence. \citet{Sutskever2014} found that using two separate RNN can be very good realization of the sequence-to-sequence prediction. In our model, the RNN we chose is GRU (Gate Recurrent Unit) invented by \citet{Cho201417} Such architecture can handle variable-length inputs corresponding to variable-frames-counts of different angiography videos. One RNN is for the encoder that receives variable-length sequence inputs (max length×batch size) and transfers the inputs into a fix length vector (max length×batch size×hidden size). The input to each GRU module is a sequence element, the output is an output vector and a hidden state (n layers×batch size×hidden size). The other RNN is the decoder, its each GRU module takes a sequence element and the fixed length vector from encoder as input and outputs a prediction of the next sequence element (batch size×class counts) and the hidden state to be passed to its next GRU (n layers×batch size×hidden size). The overall Encoder-Decoder GRU architecture is illustrated as Figure \ref{RNN}. 

\section{Results}

\begin{figure}[H]
\centering
\includegraphics[scale=0.45]{"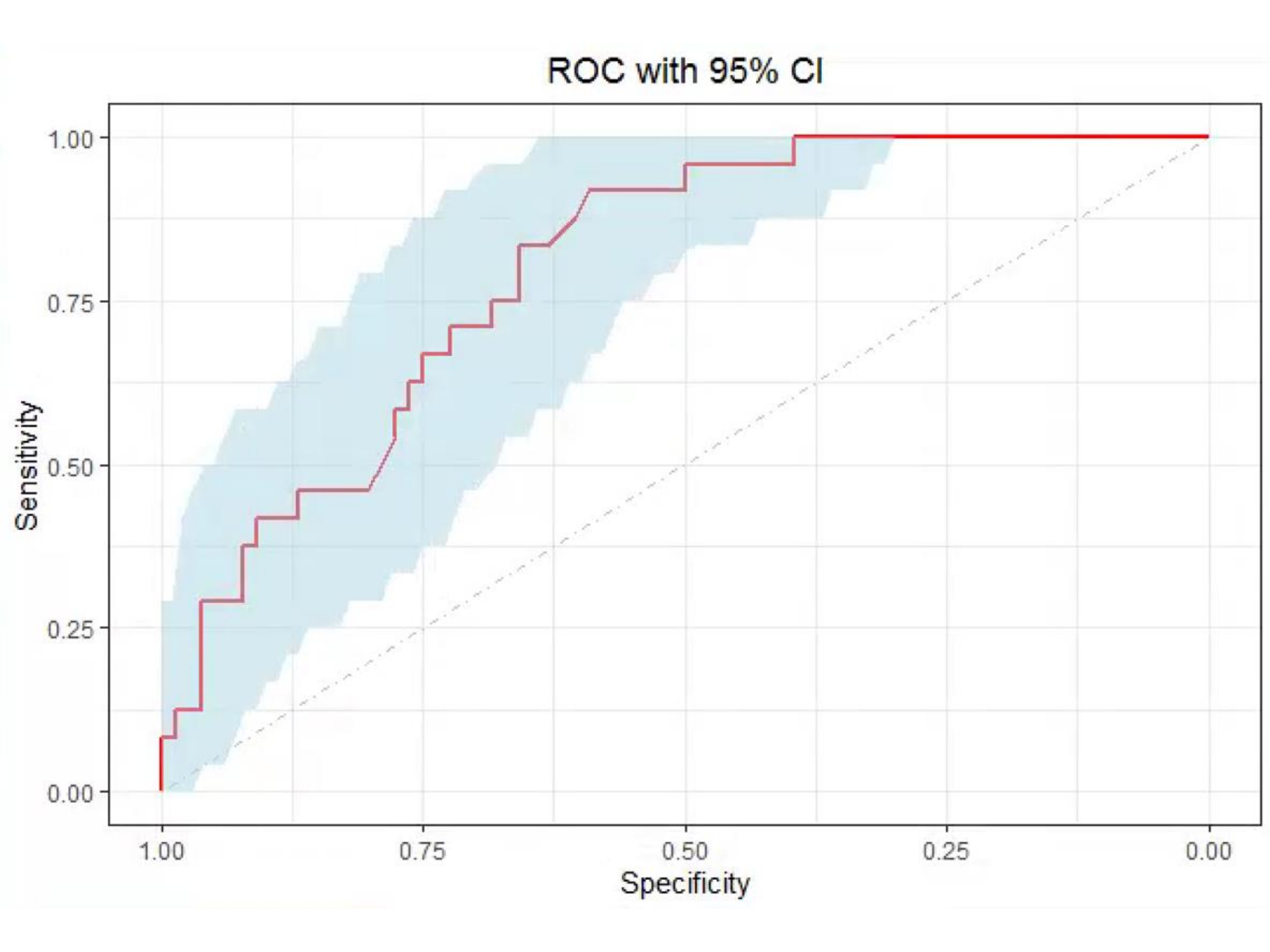"}
\caption{\textbf{Overall ROC Curve}}
\label{ROC}
\end{figure}

\begin{figure}[H]
\centering
\includegraphics[scale=0.45]{"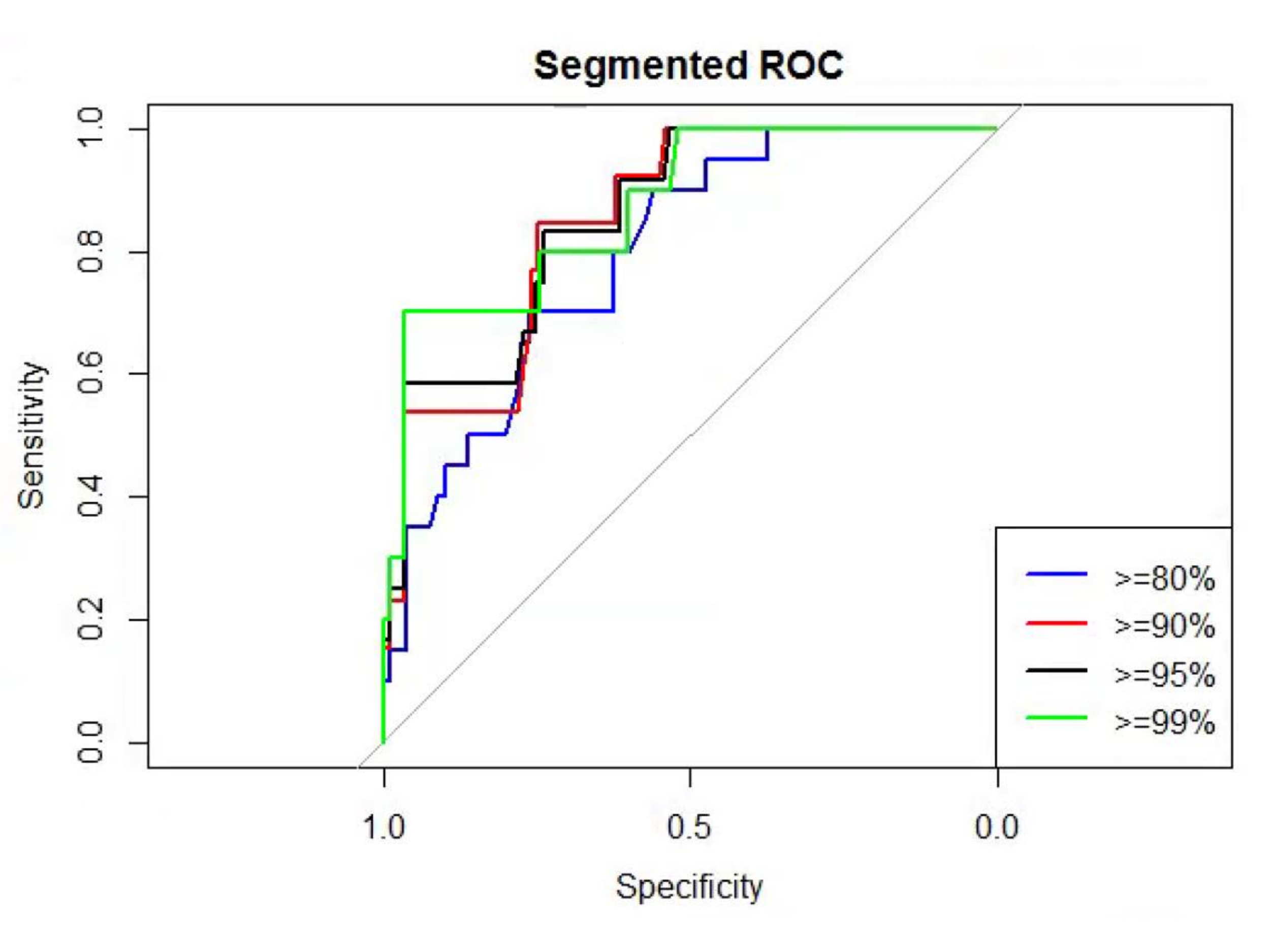"}
\caption{\textbf{Segmented ROC Curve}}
\label{segmented_ROC}
\end{figure}

We train the Encoder-Decoder GRU model described in Section\ref{Step 3: Learning} to gain the capability to recover the disarranged frames. The training set consists of 1320 Right Coronary Artery(RCA) videos and the test set consists of 100 RCA videos. The average model prediction accuracy, the corresponding average precision score and AUC is listed in Table \ref{results}, Figure \ref{ROC}, Figure \ref{segmented_ROC}.

\begin{table}[H]
	\centering
	\caption{Overall and Segmented Results}
	\label{results}
	\setlength{\tabcolsep}{1mm}{
	\begin{tabular}{p{2cm}p{3.8cm}p{3.8cm}p{2cm}}\hline 
	Stenosis & Average Test Accuracy & Average Precision Score & AUC\\
	\hline 
	\hline 
	all & 0.84 & 0.93 & 0.8\\
	\hline 
	$\geq$ 80\% & - & 0.94 & 0.79\\
	\hline 
	$\geq$ 90\% & - & 0.97 & 0.85\\
	\hline 
	$\geq$ 95\% & - & 0.98 & 0.86\\
	\hline 
	$\geq$ 99\% & - & 0.98 & 0.87\\
	\hline 
	\end{tabular}}
	
\end{table}

To highlight, the average precision score for stenosis $\geq$ 99$\%$ is up to 0.98, the corresponding AUC is up to 0.87. The results validate the intuition. Since the severe stenosis cause occlusion, hence the contrast agent injection process is blocked and similar adjacent frames are generated which increase the Reorder Difficulty significantly. Hence, the model is less able to distinguish and recover the disorder and hence generate significantly low prediction accuracy. The average prediction accuracy between positive and negative samples is most significantly different in this group so that the classification performance is enhanced most. 

\begin{figure}[H]
\centering
\includegraphics[scale=0.45]{"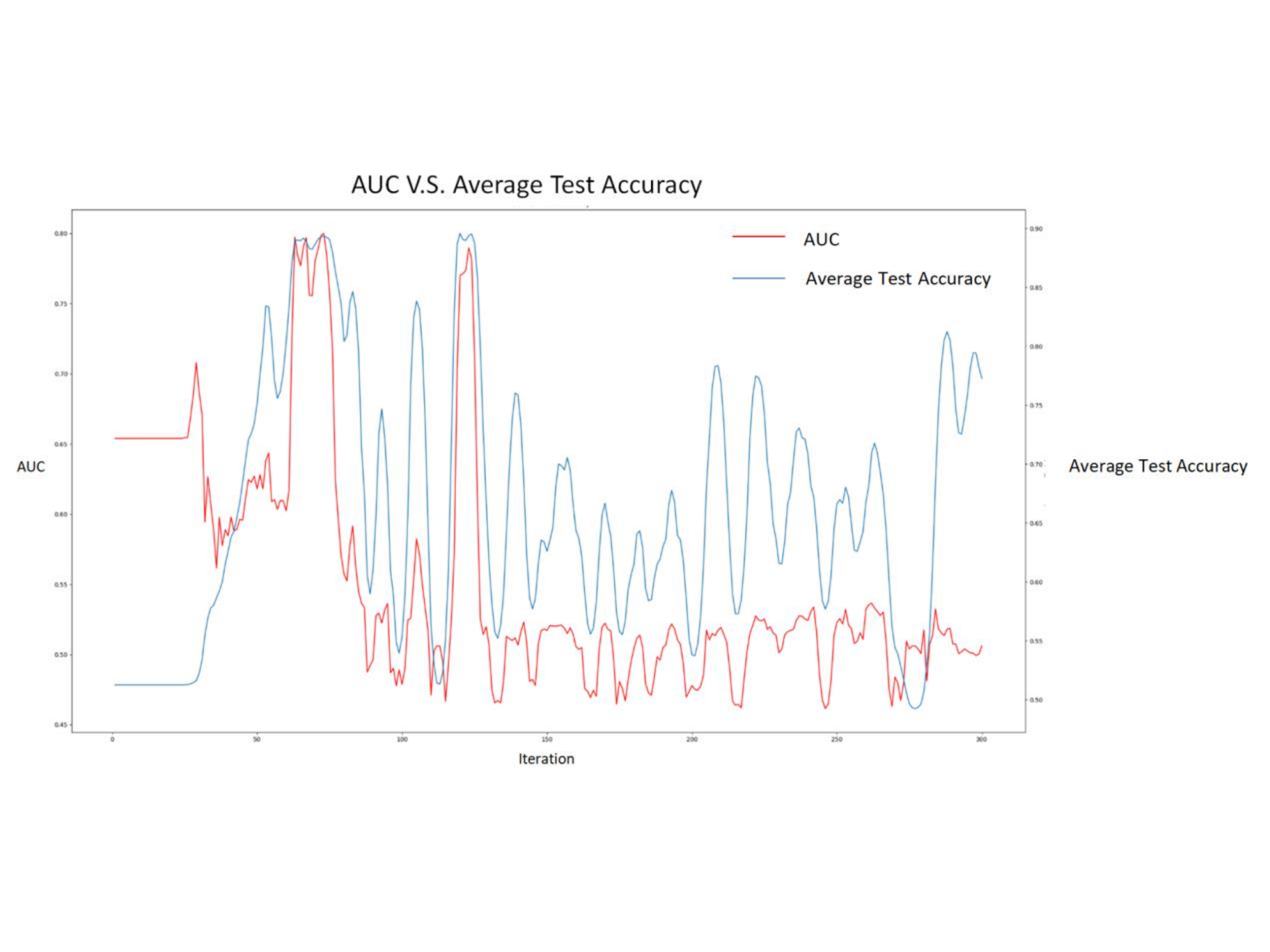"}
\caption{\textbf{Average Test Accuracy V.S. AUC}}
\label{trend}
\end{figure}

From the iteration process data of Average Test Accuracy and AUC trends (Figure \ref{trend}), we can see that they reach peak simultaneously. Generally, the higher average test accuracy is, the higher AUC it achieves. In our study, the highest average test accuracy is 0.84, leaving huge potential for the future work to improve the AUC via improving average test accuracy. 

Table \ref{comparison} is the performance comparison of several typical researches related to automatic recognition of coronary artery disease and ours. \citet{Moon2021} achieves the highest AUC of 0.92. However, their method is a supervised learning method which need laborious efforts to label engineering, so their sample size is severely constrained due to the huge labeling efforts needed. Besides, their method needs certain amounts of data preprocessing efforts such as segmentation \citep{Zana2001,Huang2018,Yang2019} and centerline extraction\citep{Gulsun2016,Wolterink2019}. \citet{Cho2019} study is totally based on traditional machine learning and needs heavy feature learning efforts. Our methodology, without any typical data preprocessing, no need to label the data, even without convolution neural network, achieves overall average precision score of 0.93 and AUC of 0.8. The highest segmented average precision score reaches 0.98 and AUC reaches 0.87.

\begin{sidewaystable}[htbp]\scriptsize
	\centering
	\caption{Performance Comparison}
   \label{comparison}
	\setlength{\tabcolsep}{0.7mm}{
	\begin{tabular}{|p{2.3cm}<{\centering}|p{1.8cm}<{\centering}|p{1cm}<{\centering}|p{1.5cm}<{\centering}|p{1cm}<{\centering}|p{1.8cm}<{\centering}|p{2.9cm}<{\centering}|p{2.9cm}<{\centering}|p{3.2cm}<{\centering}|p{2.7cm}<{\centering}|}
	\hline 
	Study & Sample Size & Device & Accuracy & AUC & Need Label & Convolution Used & Segmentation Used & Centerline Extrac. Used & Feature Engi. Used\tabularnewline
	\hline 
	\hline 
	\citet{Cho2019} & 1501 & CCTA &  & 0.87 & YES & YES & NO & NO & YES\tabularnewline
	\hline 
	\citet{Moon2021} & 452 & CAG &  & 0.92 & YES & YES & YES & YES & NO\tabularnewline
	\hline 
	\citet{Zreik2019} & 163 & CCTA & 0.8 &  & YES & YES & NO & YES & NO\tabularnewline
	\hline 
	\citet{Kumamaru2020} & 1052 & CCTA & 0.76 & 0.78 & YES & YES & NO & NO & NO\tabularnewline
	\hline 
	Proposed & 1320 & CAG &  & 0.8 & NO & NO & NO & NO & NO\tabularnewline
	\hline 
	\end{tabular}}
\end{sidewaystable}

\section{Discussion}

There are many cardiac related researches based on machine learning, especially deep learning. \citet{Isensee2020} use 2D and 3D CNN to extract the time series features of sequential images and the deep learning and random forest model are integrated to predict heart diseases. \citet{Xu2017} construct a parallel neural network model of CNN and RNN (LSTM) to predict the lesion area of myocardial infarction. \citet{Lin2020} infer the probability of coronary heart disease via facial recognition. 

\subsection{Dynamic Based}

In terms of coronary heart disease diagnostics, as far as we know, all the existing researches are static based on anatomy. That is, the analysis is totally based on separate medical image typically generated from Coronary Computed Tomography Angioplasty (CCTA). \citet{Zreik2019} use semi-supervised learning based on CCTA images to evaluate if the patient needs further invasive coronary angiography. \citet{Fischer2020} use RNN to predict the calcification fraction of coronary vessels based on CCTA. \citet{Shadmi2018} use deep CNN to calculate the calcification fraction based on CCTA. \citet{Zreik2018} use recurrenct CNN to classify coronary stenosis also based on CCTA. There are few deep learning research based on coronary angiography. \citet{Moon2021} use CNN to achieve the automatic recognition of coronary lesion based on coronary angiography. However, the preprocessing technique of their method is to select key frames instead of utilize the whole angiography video. Besides, their method is a purely supervised learning which needs data manual label efforts a lot. Analysis based on static image corresponds to anatomy diagnostic while a stenosis found in a static image doesn't necessarily mean ischemia.\citep{Tonino2019} Our methodology is based on analysis of the whole coronary angiography video instead of separate image which is theoretically dynamic in a functionally perspective. 

\subsection{Label Free}

There are many mature annotation libraries in image recognition, the most famous one is ImageNet\citep{Deng2009}, and there are some other annotation libraries in video recognition such as Kinetics\citep{Kay2017}. These databases cover general scenarios. However, in the field of medical imaging, especially coronary angiography, there is no mature annotation library developed. For studies using deep learning techniques in this field, researchers need to complete the label engineering work by themselves which is highly costly since the annotation work must be finished by professional doctors.\citep{Cho2019,Zreik2018,Zreik2019,Moon2021,Fischer2020,Kumamaru2020} This makes the development of deep learning, especially supervised learning in medical field greatly limited. According to statistics by May 2020 from OECD (Organization for Economic Cooperation and Development), the average annual income of a specialist in hospitals of United States is 350300 dollars, and for a GP is 242400 dollars. That is to say, a specialist earns up to \$168 an hour and a GP earns up to \$116 an hour, even without taking into account national holidays. Doctors estimate that it takes about 15 to 20 minutes to fully evaluate a coronary angiography, so even if a more junior GP to do the annotation, the cost is as high as \$30 for each coronary angiography. For deep learning, a lot of labeled angiographies are needed. For example, in this paper, 1320 videos of RCA angiography were used for training and 100 for testing, and the cost of manual labeling alone was at least as high as \$44080, not to mention some larger training program. This may be an important bottleneck restricting deep learning in coronary angiography. 

\begin{figure}[H]
\centering
\includegraphics[scale=0.45]{"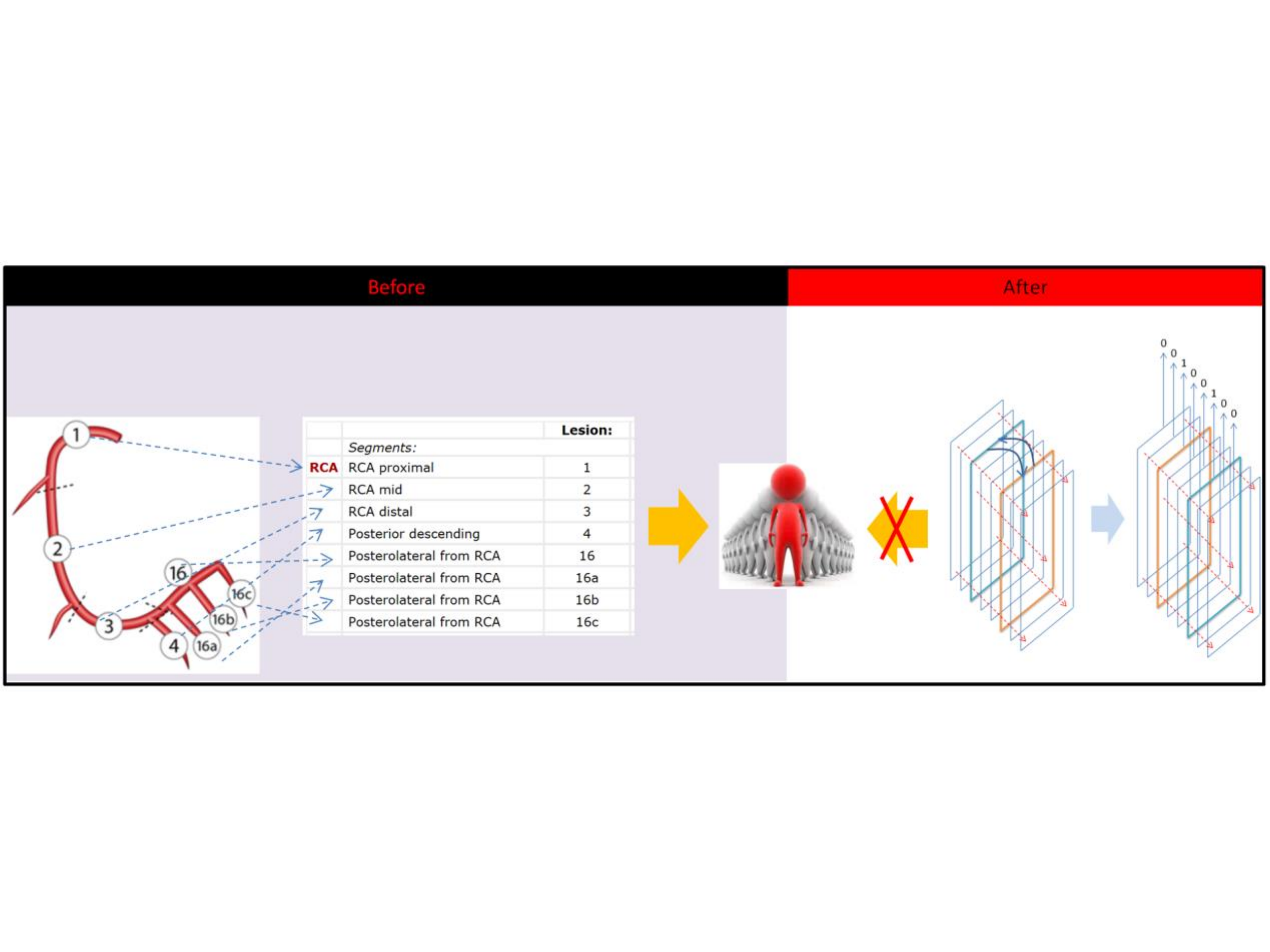"}
\caption{\textbf{The Laborious Efforts Saved Before and After.}From the Syntax Score scoring process, in the right coronary artery(RCA), there are 8 positions to potentially have lesions. Hence there are 256 combination possibilities to be diagnosed as an abnormal vessel. That is to say, to label the RCA, doctors need to exclude the 256 possibilities so as to make a final decision of a normal vessel. Hence the labeling efforts are laborious. On the other hand, the methodology in this study is 100$\%$ label free and can be generalized very easily.}
\label{label_saved}
\end{figure}

This methodology links Sequence Intensity with Coronary Artery Stenosis Status and finally demonstrates a significant relationship between them. However, there is no need to manual labeling at all so that the training cost is greatly saved and large-scale training becomes possible.(Figure \ref{label_saved})

\subsection{Potential of the Methodology}

From the performance of final results, we can see that the methodology still has great potential to improve. The test accuracy of this methodology is 0.84, the corresponding AUC can reach 0.8. As you can see from the iterations in the Figure \ref{trend}, there is a significant positive correlation between test accuracy and AUC. If test accuracy is further improved, AUC is possible to be further increased. 

\begin{figure}[H]
\centering
\includegraphics[scale=0.45]{"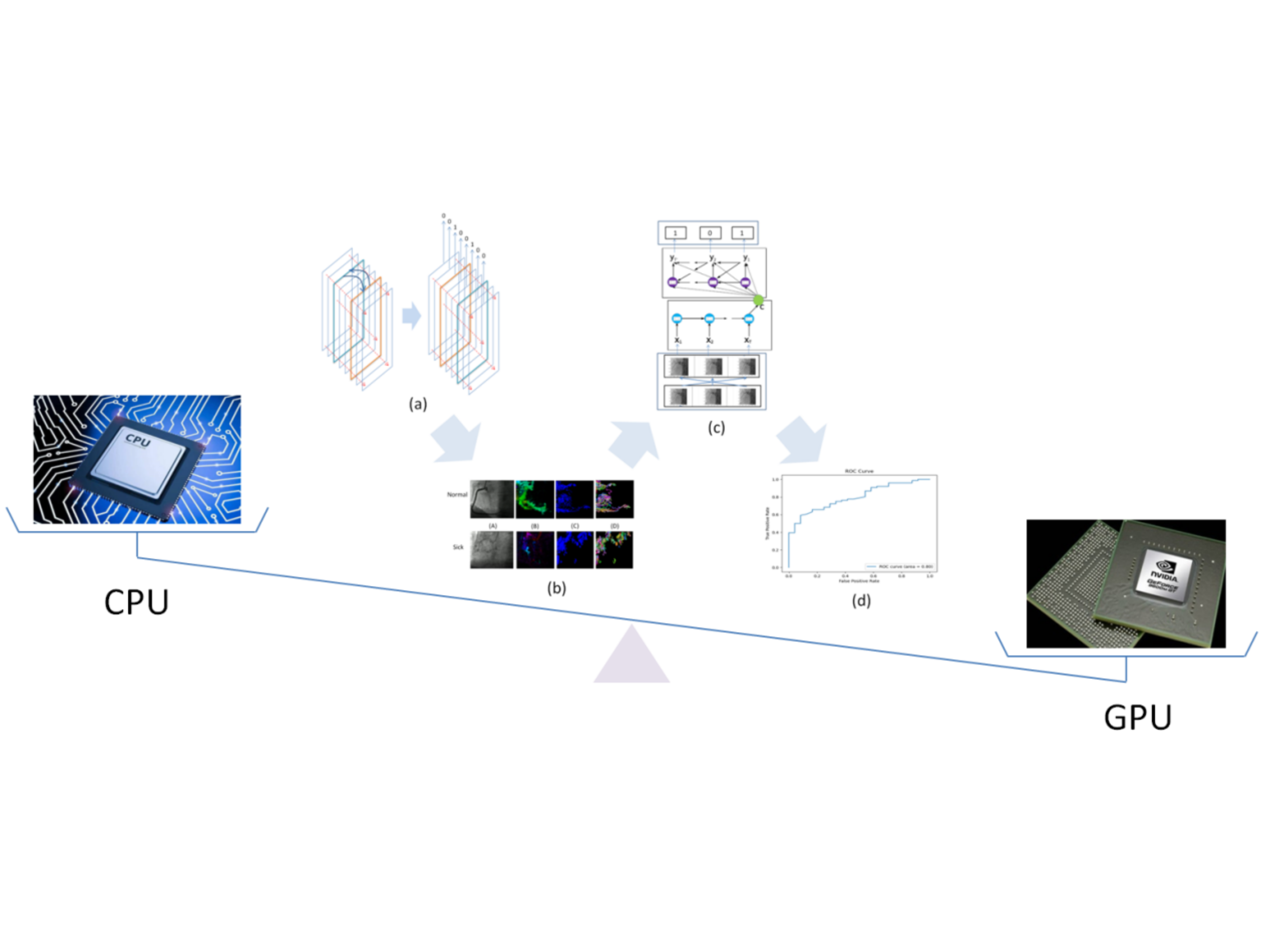"}
\caption{\textbf{Potential Improvements.} The current implementation is GPU-off. The potential to improve the performance of the methodology is huge if GPU is implemented.}
\label{GPU}
\end{figure}

The methodology uses optical flow technology to represent sequence feature between frames which is an idea of dimension reduction. The dimension of the model input reduces by 2621.44 times which is very considerable while the performance of the methodology is maintained in a good level. However, dimension reduction will eventually bring about the loss of information and indirectly affect the final test accuracy and AUC. If we can enhance the parallel computing power of computers, such as using GPU instead of CPU (Figure \ref{GPU}) to design and implement the methodology, it will improve the accuracy of model prediction to a certain extent, thus improving the final evaluation of lesions.

\subsection{Software Implementation}

\begin{figure}[H]
\centering
\includegraphics[scale=0.45]{"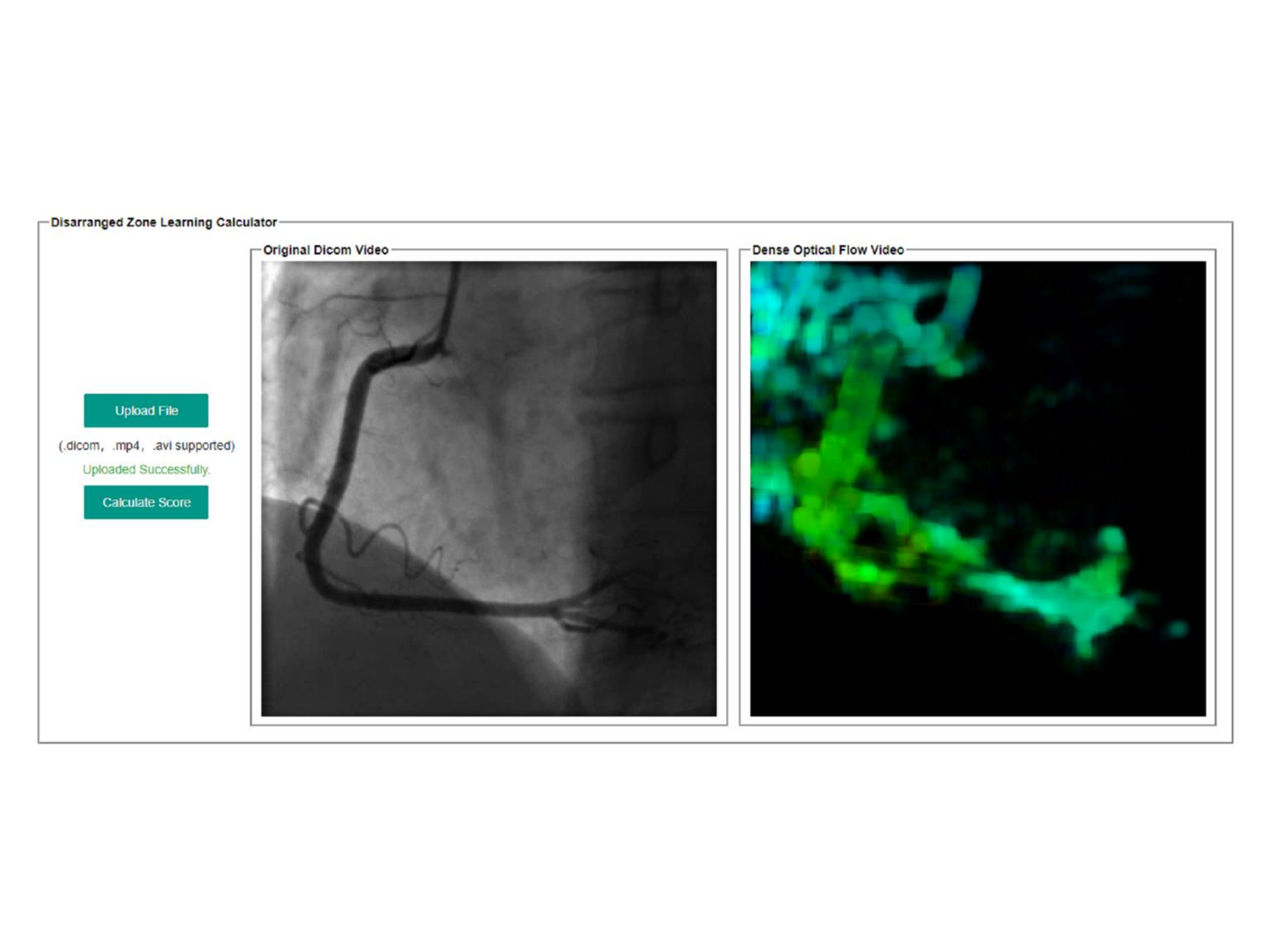"}
\caption{\textbf{Software Interface}}
\label{software_interface}
\end{figure}

\begin{figure}[H]
\centering
\includegraphics[scale=0.45]{"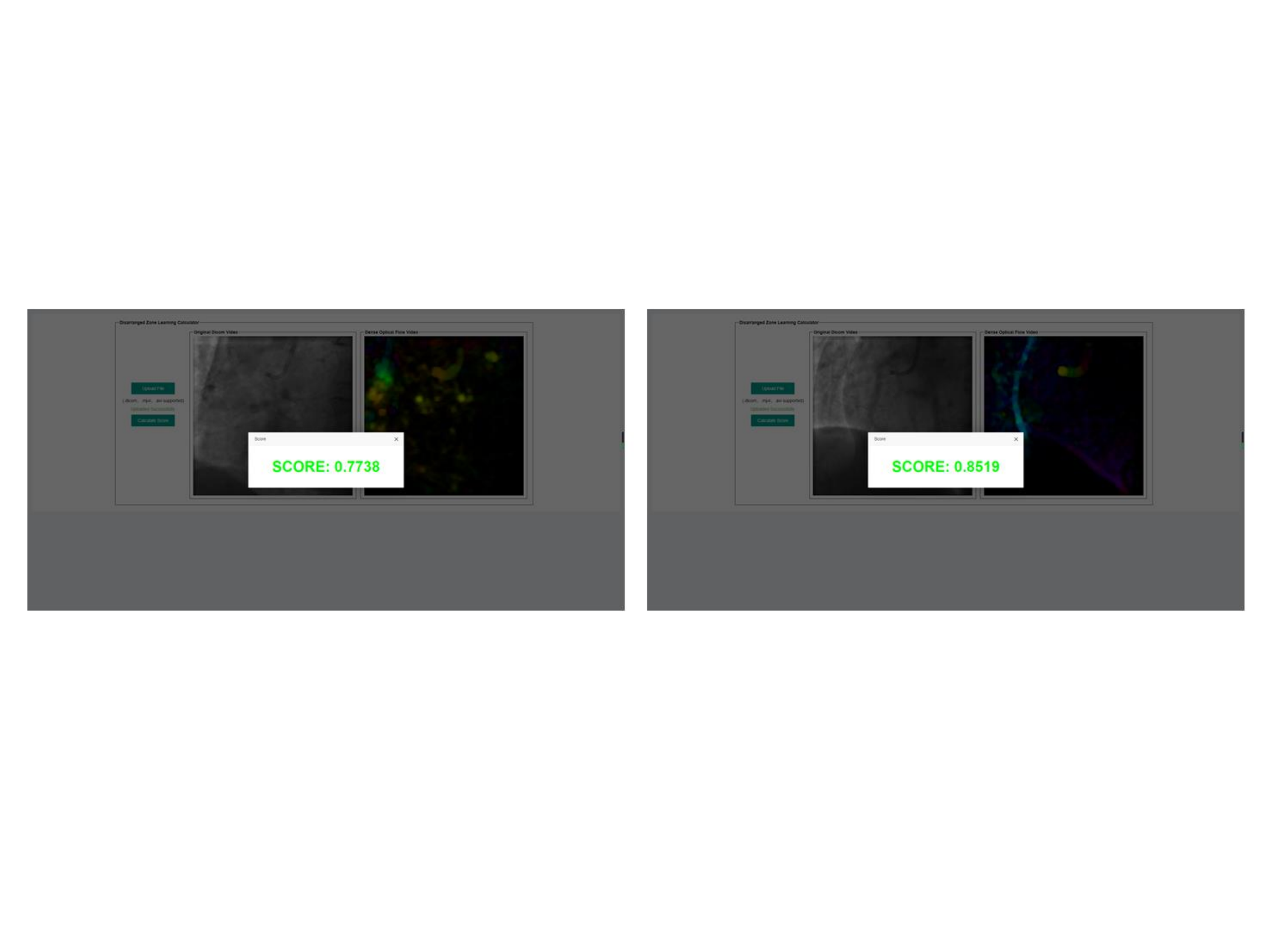"}
\caption{\textbf{Scores Generation.} The left RCA coronary angiography uploaded in our example is an abnormal video with occlusion while the right RCA coronary angiography uploaded in our example is a normal video. The final SCOREs are calculated: the occlusion video gets 0.77 and the normal video gets 0.85.}
\label{software_results}
\end{figure}

We implemented DZL methodology and developed a software to generate SCOREs for any right coronary angiography in DICOM, AVI and MP4 forms. The software interface is as Figure \ref{software_interface}. From Figure \ref{software_results}, the left side RCA coronary angiography uploaded in our example is an abnormal video with occlusion while the right side RCA coronary angiography uploaded in our example is a normal segment. The final SCOREs are calculated as Figure \ref{software_results}, the occlusion video gets 0.77 and the normal video gets 0.85. You can find the DZL calculator on http://47.103.11.61/DisarrangedZoneLearningCalculator/

\subsection{Conclusion}

This study originally proposed a novel unsupervised and dynamic methodology (DZL) to automatically recognize coronary artery stenosis. DZL is labe-free, with no precondition for inputted coronary angiography and takes advantage of the whole coronary angiography video information. DZL is easy to be generalized and migrated. Finally, we developed a software to implement the DZL for clinical practice. 

\section{Limitation}

In this study, only the Right Coronary Artery (RCA) Coronary Angiography data is used to train and test the DZL. However, the methodology is general and not constrained to any coronary artery parts. It can also be applied to Left Artery Coronary (LCA). Future work can further validate, improve and extend the methodology. As illustrated in Table \ref{comparison}, in this study, we skip all the traditional data preprocessing procedures in order to demonstrate the pure power of our methodology, however, it also constrained the performance of the methodology. Future work can gradually add the typical preprocessing procedures such as segmentation, centerline extraction etc..

\bibliography{mybibfile}

\end{document}